\documentstyle[11pt,paspconf]{article}
\input{psfig.tex}

\begin{document}

\title{Future Cosmic Microwave Background Constraints to the
    Baryon Density}

\author{M. Kamionkowski$^1$, G. Jungman$^2$,
A. Kosowsky$^{3,4}$, and D. N. Spergel$^{5,6}$}

\affil{$^1$Department of Physics, Columbia University, New York, NY 10027}

\affil{$^2$Department of Physics, Syracuse University, Syracuse, NY 13244}

\affil{$^3$Harvard-Smithsonian Center for Astrophysics,
Cambridge, MA~~02138}

\affil{$^4$Department of Physics, Harvard University, Cambridge,
MA~~02138}

\affil{$^5$Department of Astrophysical Sciences, Princeton
University, Princeton, NJ 08544}

\affil{$^6$Department of Astronomy, University of Maryland,
College Park, MD 20742}




\begin{abstract}
We discuss what can be learned about the baryon
density from an all-sky map of the cosmic microwave
background (CMB) with sub-degree angular resolution.  With only
minimal assumptions about the primordial spectrum of density
perturbations and the values of other cosmological parameters,
such a CMB map should be able to distinguish between a Universe
with a baryon density near 0.1 and a baryon-dominated Universe.
With additional reasonable assumptions, it is conceivable that
such measurements will constrain the baryon density to an
accuracy similar to that obtained from BBN calculations.
\end{abstract}


\keywords{cosmic microwave background, baryon density, big-bang
nucleosynthesis, dark matter, early-Universe cosmology,
observational cosmology}



The current range for the baryon-to-photon ratio allowed by
big-bang nucleosynthesis (BBN) is $0.0075 \la \Omega_b h^2 \la
0.024$ (Copi et al.\ 1995).  This gives
$\Omega_b\la0.1$ for the range
of acceptable values of $h$, which implies that if $\Omega=1$, as
suggested by inflationary theory (or even if $\Omega\ga0.3$ as
suggested by cluster dynamics), then the bulk of the mass in the
Universe must be nonbaryonic.  On the other hand, X-ray--cluster
measurements might be suggesting that the observed baryon
density is too high to be consistent with BBN (see, e.g., Felten
\& Steigman 1995 and references therein); this becomes
especially intriguing given the recent measurement of a large
primordial deuterium abundance in quasar absorption spectra
(Hogan \& Ruger 1995).  The range in the
BBN prediction can be traced primarily to uncertainties in the
primordial elemental abundances.  There is, of course, also some
question as to whether the X-ray--cluster measurements actually
probe the universal baryon density.  For these reasons and more,
it would clearly be desirable to have an independent measurement
of $\Omega_b h^2$.

Here, we evaluate the precision with which the baryon-to-photon
ratio, $\Omega_b h^2$, can be determined with high-resolution
CMB maps (Bennett et al.\ 1995; Janssen et al.\ 1995; Bouchet et
al.\ 1995).  We work within
the context of models with adiabatic primordial density
perturbations, although similar arguments apply to isocurvature
models as well, and we expect the power spectrum to distinguish
clearly the two classes of models (Crittenden \& Turok 1995).
(More details may be found in Jungman et al.\ 1995a,b.)

A given cosmological theory makes a statistical prediction
about the distribution of CMB temperature fluctuations,
expressed by the angular power spectrum
\begin{equation}
C(\theta) \equiv \left\langle {[\Delta T({\bf\hat m})/ T_0]}
                         {[\Delta T({\bf\hat n})/ T_0]}
			 \right\rangle_{ {\bf\hat m}\cdot{\bf\hat n} =
			 \cos\theta}
          \equiv \sum_\ell (2\ell+1) C_\ell
                     P_\ell(\cos\theta)/(4\pi),
\label{powerspectrum}
\end{equation}
where $\Delta T({\bf\hat n})/T_0$ is the fractional temperature
perturbation in the direction $\bf\hat n$, $P_\ell$ are the Legendre
polynomials, and the brackets represent an ensemble average over
all angles and observer positions.  Since we can observe from
only a single location in the Universe, the observed multipole
moments $C_\ell^{\rm obs}$ will be distributed about the mean
value $C_\ell$ with a ``cosmic variance'' $\sigma_\ell \simeq
\sqrt{2/(2\ell+1)}C_\ell$.

We consider an experiment which maps a fraction $f_{\rm sky}$
of the sky with a gaussian beam with full width at half maximum
$\theta_{\rm fwhm}$ and a pixel noise
$\sigma_{\rm pix} = s/\sqrt{t_{\rm pix}}$, where $s$ is the detector
sensitivity and $t_{\rm pix}$ is the time spent observing each
$\theta_{\rm fwhm}\times\theta_{\rm fwhm}$ pixel. We adopt the
inverse weight per solid angle,
$w^{-1}\equiv (\sigma_{\rm pix}\theta_{\rm fwhm}/T_0)^2$,
as a measure of noise that is pixel-size independent (Knox 1995).
Current state-of-the-art detectors achieve sensitivities of
$s=200\,\mu {\rm K}\,\sqrt{\rm sec}$, corresponding to an inverse
weight of $w^{-1}\simeq 2\times 10^{-15}$ for a one-year experiment.
Realistically, however, foregrounds and other systematic effects may
increase the noise level; conservatively, $w^{-1}$ will likely fall
in the range $(0.9-4)\,\times\,10^{-14}$.  Treating the pixel
noise as gaussian and ignoring any correlations between pixels,
the $C_\ell^{\rm obs}$ will be distributed about the $C_\ell$
with a standard error
\begin{equation}
\sigma_\ell = \left[{(2\ell +1)f_{\rm sky}/2}\right]^{-1/2}
              \left[C_l + (w f_{\rm sky})^{-1}
	      e^{\ell^2\sigma_b^2}\right],
\label{variance}
\end{equation}
where $\sigma_b=7.4\times10^{-3} (\theta_{\rm fwhm}/1^\circ)$.

Given a spectrum of primordial density perturbations, the $C_\ell$
are obtained by solving the coupled equations for the evolution of
perturbations to the spacetime metric and perturbations to the
phase-space densities of all particle species in the Universe.
We consider models with initial adiabatic density perturbations
filled with photons, neutrinos, baryons, and collisionless dark
matter; this includes all inflation-based models.
The CMB power spectrum depends upon many parameters.  Here,
we include the following set: the total density $\Omega$;
the Hubble constant, $H_0 = 100\;h\,{\rm km\,sec^{-1}\,Mpc^{-1}}$;
the density of baryons in units of the critical density, $\Omega_b h^2$;
the cosmological constant in units of the critical density, $\Lambda$;
the power-law indices of the initial scalar- and tensor-perturbation
spectra, $n_S$ and $n_T$; the amplitudes of the scalar and tensor
spectra, parameterized by $Q$, the total CMB quadrupole moment, and
$r=Q_T^2/Q_S^2$, the ratio of the squares of the tensor and
scalar contributions to the quadrupole moment;
the optical depth to the surface of last scatter, $\tau$; the
deviation from scale invariance of the scalar perturbations,
$\alpha\equiv dn/d\ln k$; and the effective number of
light-neutrino species at decoupling, $N_\nu$.
Thus for any given set of cosmological parameters,
${\bf s}=\{\Omega,\Omega_b h^2,h,n_S,\Lambda,r,n_T,\alpha,
\tau,Q,N_\nu\}$, we can calculate the mean multipole moments
$C_\ell({\bf s})$.

We now wish to determine the precision with which CMB maps will
be able to determine $\Omega_b h^2$ without making any
assumptions about the values of the other undetermined
parameters.  The answer will
depend on the measurement errors $\sigma_l$, and on
the underlying cosmological theory.  If the actual parameters
describing the Universe are ${\bf s}_0$, then
the probability distribution for observing a CMB power spectrum which
is best fit by the parameters ${\bf s}$ is
$P({\bf s}) \propto \exp\left\{ -{1\over 2}({\bf s}-{\bf s}_0)
\cdot [\alpha] \cdot({\bf s}-{\bf s}_0)\right\}$
where the curvature matrix $[\alpha]$ is given approximately by
\begin{equation}
\alpha_{ij} = \sum_\ell {1\over\sigma_\ell^2}
              \left[{\partial C_\ell({\bf s}_0)\over\partial s_i}
                    {\partial C_\ell({\bf s}_0)\over\partial s_j}\right]
\label{curvature}
\end{equation}
with $\sigma_\ell$ as given in Eq.~(\ref{variance}).
The covariance matrix $[{\cal C}] = [\alpha]^{-1}$ is an estimate of the
standard errors that would be obtained from a maximum-likelihood
fit to data: the error in measuring the parameter $s_i$ (obtained by
integrating over all the other parameters) is
approximately ${\cal C}_{ii}^{1/2}$. If some of the parameters
are known, then the covariance matrix for the others is
determined by inverting the submatrix of the
undetermined parameters.

\begin{figure}
\centerline{\psfig{figure=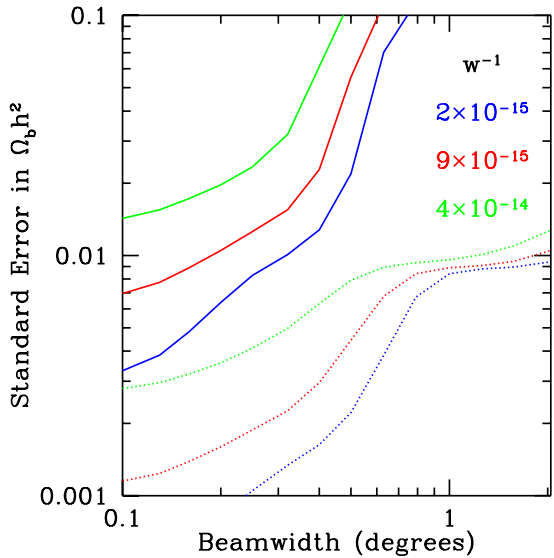,height=3in}}
\caption{The standard error on $\Omega_b h^2$.}
\label{Omegafigure}
\end{figure}

Fig.~\ref{Omegafigure}\ displays the standard error in $\Omega_b
h^2$ as a function
of the beam width $\theta_{\rm fwhm}$ for different noise levels
and for $f_{\rm sky}=1$.
The underlying model assumed here for the purpose of
illustration is ``standard CDM,'' given by ${\bf
s}=\{1,0.01,0.5,1,0,0,0,0,0,Q_{\rm COBE},3\}$, where $Q_{\rm
COBE}=20\,\mu$K is the COBE normalization (G\'orski et al.\ 1994).
The solid curves show the ${\cal C}_{\Omega_b h^2,\Omega_b
h^2}^{1/2}$ obtained by
inversion of the full $11\times11$ curvature matrix $[\alpha]$ for
$w^{-1}=2\times10^{-15}$, $9\times10^{-15}$, and $4\times
10^{-14}$.  These are the
sensitivities that can be attained at the given noise levels
with the assumption of uniform priors (that is, including {\it
no} information about any parameter values from other observations).
The dotted curves show the ${\cal C} _{\Omega_b h^2,\Omega_b h^2}^{1/2}$
obtained by inversion of the $\Omega_b h^2$-$Q$ submatrix of
$[\alpha]$; this is the
error in $\Omega_b h^2$ that could be obtained if all other
parameters except the normalization were fixed, either from
other observations or by assumption.  Realistically,
the precision obtained will fall somewhere between these
two sets of curves.  The results for a mapping experiment which
covers only a fraction $f_{\rm sky}$ of the sky can be obtained
by replacing $w \rightarrow w f_{\rm sky}$ and scaling by
$f_{\rm sky}^{-1/2}$ [c.f., Eq.~(2)].

The implications of CMB maps for the
baryon density depend quite sensitively on the experiment.  As
long as $\theta_{\rm fwhm}\la0.5$, the CMB should (with minimal
assumptions) at least be
able to rule out a baryon-dominated Universe ($\Omega_b\ga0.3$)
and therefore confirm the predictions of BBN.  With angular
resolutions that approach $0.1^\circ$ [which might be
achievable, for example, with a ground-based interferometry map
(Myers 1995) to complement a satellite map], a CMB map would
provide limits to the baryon-to-photon ratio that were
competitive with BBN.  Furthermore, if other parameters can be
fixed, the CMB might be able to restrict $\Omega_b h^2$ to a
small fraction of the range currently allowed by BBN.

Moreover, the CMB will also provide information on several other
parameters (Jungman et al.\ 1995a,b).  Most significantly, the
total density $\Omega$ can be determined to better than 10\%
with minimal assumptions and perhaps better than 1\%.

\acknowledgments

This work was supported in part by the D.O.E. under contracts
DEFG02-92-ER 40699 and DEFG02-85-ER 40231, by the Harvard
Society of Fellows, by the NSF under contract ASC 93-18185,
and by NASA under contract NAG5-3091 and NAGW-2448, and by NASA
under the MAP Mission Concept Study Proposal.

%


\begin{references}

\reference Bennett, C. L.\ et al.\ 1995, NASA Mission Concept
     Study

\reference Bouchet, F. R. et al.\ 1995, astro-ph/9507032

\reference Copi, C. J., et al.\ 1995, astro-ph/9508029

\reference Crittenden, R. G. \& Turok, N. 1995,
     Phys. Rev. Lett., 75, 2642

\reference Felten, J. E.\ \& Steigman, G. 1995, in
     Proc. St. Petersburg Gamow Seminar, St. Petersburg, Russia,
     12--14 September 1994, A. M. Bykov and A. Chevalier,
     Sp. Sci Rev. (Dordrecht: Kluwer)

\reference G\'orski, K. M. et al.\ 1994, \apjl, 430, L89

\reference Hogan, C. J.\ \& Ruger, A. 1995, this proceedings

\reference S. T. Myers, private communication

\reference Janssen, M. A. et al.\ 1995, NASA Mission Concept
     Study

\reference Jungman, G., Kamionkowski, M., Kosowsky, A., \&
     Spergel, D. N. 1995a, astro-ph/9507080, Phys. Rev. Lett.,
     in press

\reference Jungman, G., Kamionkowski, M., Kosowsky, A., \&
     Spergel, D. N. 1995b, in preparation

\reference Knox, L. 1995, Phys. Rev. D, 52, 4307

\end{references}
\end{document}